\documentclass[aps,pre,twocolumn,amssymb,amsfonts,floatfix,showpacs]{revtex4}

\usepackage{graphicx}% Include figure files
\usepackage{dcolumn}% Align table columns on decimal point
\usepackage{bm}% bold math
\usepackage{mathbbol}
\usepackage[dvips]{color}

\begin{document}

\title{Onset of chaos and relaxation in isolated systems of interacting spins-1/2:\\energy shell approach}

\author{L.~F. Santos}
\email{lsantos2@yu.edu}
\affiliation{Department of Physics, Yeshiva University, 245 Lexington Ave, New York, NY 10016, USA}
\author{F. Borgonovi}
\email{fausto.borgonovi@unicatt.it}
\affiliation{Dipartimento di Matematica e Fisica and Interdisciplinary Laboratories for Advanced Materials Physics,
Universit\'a Cattolica, via Musei 41, 25121 Brescia, and INFN,
Sezione di Pavia, Italy}
\author{F.~M. Izrailev}
\email{felix.izrailev@gmail.com}
\affiliation{Instituto de F{\'i}sica, Universidad Aut\'onoma de Puebla,
Apt. Postal J-48, Puebla, Pue., 72570, Mexico}
\affiliation{NSCL and Dept. of Physics and Astronomy, Michigan State
University, East Lansing, MI 48824-1321, USA}

\begin{abstract}
We study the onset of chaos and statistical relaxation in two isolated dynamical quantum systems of interacting spins-1/2, one of which is integrable and the other chaotic. Our approach to identifying the emergence of chaos is based on the level of delocalization of the eigenstates with respect to the energy shell, the latter being determined by the interaction strength between particles or quasi-particles. We also discuss how the onset of chaos may be anticipated by a careful analysis of the Hamiltonian matrices, even before diagonalization. We find that despite differences between the two models, their relaxation process following a quench is very similar and can be described analytically with a theory previously developed for systems with two-body random interactions. Our results imply that global features of statistical relaxation depend on the degree of spread of the eigenstates within the energy shell and may happen to both integrable and non-integrable systems.
\end{abstract}

\pacs{05.45.Mt,05.30.-d,05.70.Ln, 02.30.Ik}
\maketitle

\section{Introduction}

In recent years, a great deal of attention has been paid to the issue of thermalization in isolated quantum systems caused by interparticle interactions \cite{ETH,zele,ZelevinskyRep1996,FIC96,Flambaum1997b,BGIC98,I01,rigol,Fitzpatrick2011,lea01,lea,recent}. Apart from theoretical aspects, this interest has been triggered by remarkable experimental progresses in the studies of quantum systems with ultracold gases trapped in optical lattices (see, e.g., \cite{experim}).

A necessary condition for the onset of thermalization is the statistical relaxation of the system to some kind of equilibrium, which is followed by further fluctuations of the observables around their average values. In classical mechanics, as discussed in Ref.\cite{C97}, there are two mechanisms leading to the emergence of statistical behavior in dynamical (deterministic) systems.

The first scenario, known since the early days of statistical mechanics, is the thermodynamic limit in which the number of particles diverges, $N \rightarrow \infty$. In this case, the statistical description is valid even in the absence of chaos. A completely integrable system, such as the Toda-lattice, can manifest perfect statistical and thermodynamical properties for a finite, although large number of particles (practically, for $ N \gg 1$ \cite{CCPC97}). Even though there are initial conditions which correspond to solitons, they are rare and can be safely neglected in practice. This first mechanism, termed ``linear chaos" in Ref.\cite{C97}, is at the core of the foundation of statistical mechanics.

The other mechanism, which is more recent, is based on the concept of local instability of motion in phase space. The understanding is that an isolated dynamical system can behave in a statistical way even for a very small number of interacting particles, $N \geq 2$, provided the motion is strongly chaotic (see, e.g. \cite{C79,LL83}). Chaoticity does not imply ``true" randomness in the equations of motion, but a ``pseudo-randomness" (or, {\it deterministic chaos}), which depends on the number of particles and the strength of the interparticle interaction. Ergodicity is not essential here, provided the measure of initial conditions corresponding to regular motion is very small. In this case, an apparent irreversibility of motion emerges, since any weak external perturbation gives rise to non-recurrence of the initial conditions.

It should be stressed that, although the two mechanisms above are different, in both cases the time dependence of the observables can be described by an infinite number of statistically independent frequencies (see details in Ref.\cite{C97}).

In quantum systems, the notion of trajectories and thus of their local instabilities loses its meaning. Yet, it has been argued that thermalization may still happen, even if the system is finite and isolated, provided it is chaotic. Chaos at the quantum level refers to specific properties of spectra, eigenstates, and dynamics of the system. They were initially observed in quantum systems whose classical counterparts were chaotic, but were soon found also in quantum systems without a classical limit and in quantum systems with disordered potentials. Nowadays, the term {\it quantum chaos} is used in a broad sense when referring to those properties, irrespectively of the existence of a true classical limit.

After intensive investigation, the properties of {\it one-body} quantum chaos became well understood (see, e.g., Refs.~\cite{CC95,ReichlBook,S99}). In contrast, the theory of {\it many-body} chaos with respect to quantum systems of interacting Fermi or Bose particles is far from being complete. In fact, even in the classical limit, a proper analysis of
chaos becomes complicated due to the large number of interacting particles and, therefore, large dimensionality of the phase space.

Initial studies of quantum chaos in many-body systems focused on the statistics of the energy levels. But it soon became clear that crucial information is contained in the eigenstates. Typically, the eigenstates are written in the basis corresponding to non-interacting particles. This corresponds to using a picture where the total Hamiltonian of the model is separated into a sum of two terms, $H=H_0 + V$, where $H_0$ describes the non-interacting particles (in a more general context, quasi-particles), and $V$ absorbs the interparticle interactions. In nuclear physics the latter term is referred to as ``residual interaction".

The separation of the Hamiltonian into two different parts is, in fact, nothing but the mean-field (mf) approach, widely used in atomic and nuclear physics. In many cases, the choice of unperturbed mf-basis in which $H_0$ is diagonal is not well-defined (not unique). However, this choice is usually well supported physically, especially when the interaction between particles can be considered small. Examples include interactions between outer shell electrons in atoms, electrons in quantum dots, and interactions between spins.

The key point of many-body quantum chaos is that the eigenfunctions (EFs) in the mf-basis spread as the interaction between particles increases and may eventually have a very large number of contributing components. However, contrary to full random matrices, where the eigenstates are completely extended independently of the choice of basis, in isolated systems with finite-range interactions, the perturbation couples only part of the unperturbed basis states $|n\rangle$. Therefore, only a fraction of the coefficients $C_n^\alpha$ composing the full Hamiltonian eigenstates $|\alpha \rangle =\sum_{n} C^{\alpha}_{n} |n\rangle $ can be essentially different from zero. In the energy representation, this fraction constitutes the
{\it energy shell} of the system, which can be partly or fully filled by the exact eigenstates~\cite{Casati1993,Casati1996}. When the number of non-zero elements $C^{\alpha}_{n}$ is a small portion of the shell, the eigenstates are {\it localized}, while a large portion implies either {\it sparse} or {\it ergodic} states~\cite{QCC}. In ergodic eigenstates, the coefficients $C^{\alpha}_{n}$ become random variables following a Gaussian distribution around the ``envelope" defined by the energy shell. This latter scenario is used as a rigorous definition of {\it chaotic eigenstates} and occurs when the interaction exceeds a critical value~\cite{Casati1993,Casati1996,zele,ZelevinskyRep1996,FIC96,Flambaum1997b}. An example of such chaotic eigenstates was reported in Ref.\cite{C85}, where a careful analysis of experimental data for the cerium atom revealed that excited states with fixed total angular momentum and parity $J^{\pi} = 1^{+}$ are {\it random superpositions} of a restricted number of basis states. 

The energy shell is associated with the limiting form of the {\it strength function} (SF) written in the energy representation~\cite{Casati1993,Casati1996}. This function is obtained by projecting the unperturbed states onto the basis of exact eigenstates. It is also known as the  {\it local density of states} and is broadly used in nuclear and solid state physics. SF contains much information about global properties of the interactions. It has been shown, for example, that its shape changes from Breit-Wigner (Lorentzian) to Gaussian as the interparticle interaction increases~\cite{zele,ZelevinskyRep1996,FI00,Flambaum1997b,Kota1998,Kota2001}. 

When the eigenstates are chaotic and the quantum system has a well defined classical limit, the shapes of both EFs and SFs in the energy representation have classical analogs \cite{Casati1993,Casati1996,QCC}. The first matches the distribution of the projection of the phase space surface of $H$ onto $H_0$, and the second the projection of the surface of $H_0$ onto $H$. The onset of delocalization of EFs in the energy shell is then directly related to the {\it chaotization} of the system in the classical limit~\cite{QCC} and provides a tool to reveal the transition to quantum chaos even for dynamical quantum systems without a classical limit.

The emergence of chaotic eigenstates has been related to the onset of thermalization in isolated quantum many-body systems \cite{ETH,zele,ZelevinskyRep1996,FIC96,Flambaum1997b,BGIC98,I01,rigol,Fitzpatrick2011,lea01,lea}. It has been shown, for instance, that when the eigenstates become chaotic, the distribution of occupation numbers achieves standard Fermi-Dirac or Bose-Einstein forms, thus allowing for the introduction of temperature \cite{FIC96,Flambaum1997b,BGIC98,I01}. In particular, an analytic expression connecting the increase of temperature with the interaction strength and the number of particles was obtained using a two-body random matrix model~\cite{Flambaum1997b}. Therefore, the interparticle interaction plays the role of a heat bath for the isolated system. Another important aspect is that since the components of chaotic eigenstates can be treated as random variables, the eigenstates close in energy are statistically similar. This fact is at the heart of the so-called {\it Eigenstate Thermalization Hypothesis} (ETH) \cite{ETH} and has been employed to justify the agreement between the expectation values of few-body observables and the predictions from the  microcanonical ensemble \cite{ETH,rigol,lea01,lea}.

%%%%%%%%%%%%%%%%%%%%%%%%%%%%%%%%%%%%%%%%%%%%%%%%%%%%%%%%%%%%%%%%%%%%%%%%%%%%%%%%%%%%%%%%%%%%

The aim of the present work is to analyze the emergence of statistical properties in isolated quantum many-body systems. We consider two dynamical models of interacting spins-1/2; one is integrable for any value of the perturbation and the other undergo a transition to chaos. Our approach is based on the concept of the energy shell, in which the eigenstates undergo a transition from localized or sparse to delocalized and random. Strictly speaking, chaotic eigenstates filling completely the energy shell appear only for the nonintegrable model. However, even for the integrable system, chaotic-like eigenstates, where a large number of mf-basis contributes to the state, may be found in the limit of strong interaction. We demonstrate that the critical strength of the interaction above which the eigenstates may be considered chaotic-like corresponds to the point where the shape of SF becomes Gaussian.

We show that in comparison with the chaotic model, the lack of ergodicity of EFs in the integrable system leads to larger fluctuations of the delocalization measures and for the overlaps between neighboring eigenstates. This coincides with recent results obtained for bosonic and fermionic systems~\cite{lea01,lea}. In the spirit of ETH, these findings were used to explain the better agreement between eigenstate expectation values of few-body observables and thermal averages for systems in the chaotic domain.

Despite differences in some static properties, the relaxation process for both models after a quench is found to be very similar, as inferred from the study of the time dependence of the Shannon entropy for initial states corresponding to mf-basis states. Our numerical data agree very well with analytical predictions developed for two-body random matrices~\cite{Flambaum2001b}, when the interaction strength is strong. In this case, the entropy shows a linear growth before reaching complete relaxation. Crucial for this behavior is that the eigenstates are delocalized (although not necessarily ergodic) in the energy shell, which may occur even when the system is integrable.

We also discuss how one can predict the onset of chaotic-like eigenstates by analyzing the structure of the Hamiltonian matrices without resorting to their diagonalization. Remarkably, the estimates coincide very closely with the critical values obtained from energy level statistics and the shapes of SF and EF.

The paper is organized as follows. Section \ref{Sec:model} describes the models studied, their symmetries, and the structure of the Hamiltonian matrices. Section \ref{Sec:chaos} analyzes the fluctuations of the energy spectrum and quantifies the level of chaoticity of the system based on the level spacing distributions. Section \ref{Sec:EFs} investigates the integrable-chaos transition from the perspective of the eigenstates. We study the shape of the strength functions, the spreading of the eigenstates in the energy shell, and delocalization measures. We also propose a new signature of chaos based on correlations between neighboring eigenstates. Section \ref{Sec:Shannon} focuses on the time evolution of the Shannon entropy for both integrable and nonintegrable models aiming at identifying the conditions for statistical relaxation. Both numerical and analytical results are provided. Concluding remarks are presented in Sec.\ \ref{Sec:conclusion}.
%%%%%%%%%%%%%%%%%%%%%%%%%%%%%%%%%%%%%%%%%%%%%%%%%%%%%%%%%%%%%%%%%%%%%%%%%%%%%%%%%%%%

\section{System Model}
\label{Sec:model}

We consider isolated one-dimensional (1D) systems of interacting spins-1/2. These prototype quantum many-body systems are employed in the studies of a variety of subjects, ranging from quantum computing~\cite{Gershenfeld1997,Loss1998,Kane1998} and quantum phase transition~\cite{Coldea2010} to the transport behavior in magnetic compounds~\cite{Zotos1997,Sologubenko2000,Heidrich2004,Zotos2005,Hlubek2010,Sirker2011,Santos2011}.
The recent viability to experimentally realize such models in optical lattices~\cite{Duan2003,Trotzky2008,Simon2011,ChenARXIV} have further increased the interest in them.
In 1D, these systems may remain integrable even in the presence of interaction; while the crossover to chaos can be induced by different integrability breaking terms~\cite{Hsu1993,Avishai2002,Santos2004,Rabson2004,Kudo2005,Dukesz2009}. This particularity turns them into natural testbeds for the analysis of the integrable-chaos transition
and for comparative studies between the two regimes.

Two 1D spin-1/2 systems are investigated in this work. Model 1 has only nearest-neighbor (NN)
couplings and is integrable for any value of the interaction strength. Model 2 includes
nearest and next-nearest-neighbor (NNN) couplings, and it becomes chaotic when the
strengths of the two are comparable. Both are dynamical systems, that is they are devoid of random elements. The source of chaos in such scenarios is the complexity derived from the interparticle interactions.

\subsection{Hamiltonian}

The Hamiltonians for Model 1 and Model 2 are respectively given by

\begin{eqnarray}
&& H_1 = H_0 + \mu V_1 ,
\label{model1} \\
&& H_0 = \sum_{i=1}^{L-1} J \left(S_i^x S_{i+1}^x + S_i^y S_{i+1}^y \right) ,
\nonumber \\
&& V_1 = \sum_{i=1}^{L-1} J  S_i^z S_{i+1}^z,
\nonumber
\end{eqnarray}
and
\begin{eqnarray}
&& H_2 = H_1 + \lambda V_2 ,
\label{model2} \\
&& V_2 = \sum_{i=1}^{L-2} J\left[ \left( S_i^x S_{i+2}^x + S_i^y S_{i+2}^y
\right) + \mu S_i^z S_{i+2}^z  \right] .
\nonumber
\end{eqnarray}
Above,  $\hbar$ is set to 1, $L$ is the number of sites, and $S^{x,y,z}_i = \sigma^{x,y,z}_i/2$ are the spin operators at site $i$,  $\sigma^{x,y,z}_i$ being the Pauli matrices.
The coupling parameter $J$ determines the energy scale and is set to 1. The Zeeman splittings, caused by a static magnetic field in the $z$ direction,
are the same for all sites and are not shown in the Hamiltonians above. We refer to a spin pointing up in the $z$ direction as an excitation.

(i) In Model 1, $H_0$ corresponds to the unperturbed part of the Hamiltonian and $\mu$ is the
strength of the perturbation. The unperturbed part is known as the flip-flop term and is responsible for moving the excitations
through the chain. A system described by $H_0$ is integrable and can be mapped onto a system of noninteracting spinless fermions~\cite{Jordan1928} or hardcore bosons~\cite{Holstein1940}.
It remains integrable with the addition of the Ising interaction $V_1$, no matter how large the anisotropy parameter $\mu $ is. The total Hamiltonian $H_1$ is referred to as the XXZ Hamiltonian
and can be solved with the Bethe Ansatz~\cite{Bethe1931,Alcaraz1987,Karbach1997}. We assume $J$ and $\mu$ positive, thus favoring antiferromagnetic order.

(ii) The unperturbed part of Model 2 is the XXZ Hamiltonian. The parameter $\lambda$ refers to the ratio between the NNN
exchange, as determined by the perturbation $V_2$, and the NN couplings, characterized by $H_1$. A sufficiently large $\lambda$ leads to the onset of chaos.

With respect to symmetries, conservation of total spin in the $z$ direction, $S^z=\sum_{i=1}^{L} S_i^z$, occurs for all parameters of Hamiltonians (\ref{model1}) and (\ref{model2}).
Our analysis is thus
restricted to a particular $S^z$-subspace. In order to deal with a reasonably large sector without resorting to very large system sizes, other symmetries~\cite{Brown2008} are avoided as follows.

$\bullet$ We deal with open boundary conditions, instead of closed boundary conditions, to prevent momentum conservation.

$\bullet$ We choose subspaces filled with $L/3$ up-spins to guarantee that $S^z \neq 0$. The $S^z=0$ sector,  which appears when the chain size is even and has $L/2$ up-spins, shows invariance under a $\pi$-rotation around the $x$-axis. The dimension of the $S^z$-subspace that we consider is therefore $D_{L/3} = L!/[(L/3)!(L-L/3)!]$. Unless stated otherwise, all figures are obtained for $L=15$.

$\bullet$ We use $\mu \neq 1$ throughout to circumvent conservation of total spin, $S^2=(\sum_{i=1}^L \vec{S}_i)^2$. Different values of $\mu $ are studied for Model 1, but for Model 2, where the main interest is in the effects of the integrability breaking term $V_2$, we fix $\mu = 0.5$.

$\bullet$ Parity is not avoided. We take it into account by analyzing even and odd eigenstates separately. The dimension of each parity sector is $D_P \sim D_{L/3}/2$.

Since our numerical studies require all eigenvalues and eigenvectors of the systems,
exact full diagonalization is performed. However, as it will be clear along the text,
much information can be obtained just from the Hamiltonian matrix itself.

\subsection{Structure of the Hamiltonian matrix and strength of the perturbation}

An essential point for the study of the Hamiltonian matrix is the basis considered. In general, the choice of basis is made on physical grounds, depending on the question being addressed. In the case of the Fermi-Pasta-Ulam model, for example, one focuses on the equipartition of energy among normal modes~\cite{Berman2005}. When studying spatial localization, on the other hand, the most appropriate basis
is the coordinate basis, which in the case of lattice systems corresponds to the site-basis. For systems (\ref{model1}) and (\ref{model2}), the site-basis corresponds to arrays of spins pointing up and down in the $z$ direction.

Here, our goal is to understand the effects of the residual perturbations $V_1$ and $V_2$. They add complexity to the system, without necessarily bringing it to the chaotic domain. It becomes then essential to select a basis associated with the uncoupled particles (or quasiparticles) with which we may separate regular from complex behavior. This is the role of a mf-basis, which appears in various contexts of many-body physics. The derivation of Fermi-Dirac or Bose-Einstein distributions, for instance, requires the selection of a mf-basis. The same is true when studying the structures of nuclear and atomic systems, as well as their transition to quantum chaos. Nevertheless, there is not a well defined mathematical recipe to identifying the mf-basis; this is done based on the physical properties of the system. For the total Hamiltonians $H_1$ and $H_2$, we choose the eigenstates of  $H_0$ and $H_1$,respectively, as the unperturbed basis states, $|n\rangle $.

\begin{figure}[htb]
\includegraphics[width=0.45\textwidth]{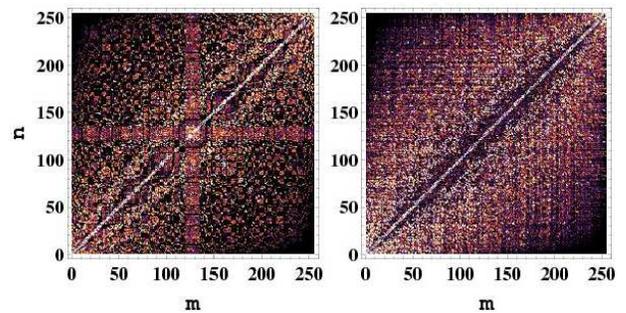}
\caption{(Color online.) Absolute values of the matrix elements of Model 1 (left panel) and Model 2 with $\lambda =0.5$ (right panel) for $L=12$ [therefore $D_P\sim 250$] and $\mu=0.5$. The mf-basis is ordered in energy. Only even states are considered. Light color indicates large values.}
\label{fig:structure}
\end{figure}

To give an idea on how to extract information from the Hamiltonian matrix, we show in
Figure~\ref{fig:structure} the density plot of the absolute values of the matrix elements for
 Model 1 (left panel) and Model 2 (right panel).
The matrices are written in the mf-basis, the latter being ordered from lowest to highest energy. Light colors indicate large values. Only elements associated with even states are shown, so no trivial symmetries are present. Both matrices have large diagonal elements and significant couplings even
between distant basis vectors. It is only far from the diagonal that the elements fade away,
as expected for realistic physical models. More zeros are found in the matrix of Model 1, which is thus more sparse than the matrix of Model 2.
Both matrices are obviously symmetric with respect to the diagonal,
since $H_{nm}=H_{mn}$. In addition to this, Model 1 shows an impressive regular structure which must be related to its integrability; various curves of high density suggest strong correlations between the matrix elements. For example, for the lines in the middle, such as $mid= 121, 122, \ldots 135$, we find that several elements, but not all, satisfy the relation $|H_{mid,1+k}|=|H_{mid,D_P-k}|$. We leave it for a future publication the interesting exercise of identifying the sources of such correlations.

\begin{figure}[htb]
\includegraphics[width=0.45\textwidth]{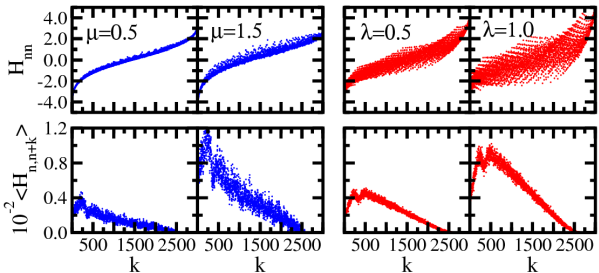}
\caption{(Color online.) Information about the matrix elements of Model 1 (two left columns) and Model 2 (two right columns). The matrices are written in the mf-basis, which is ordered from lowest to highest energy. The perturbation strength for each column is shown in the top panels. Top panels: diagonal elements. Bottom panels: average values of the absolute values of the off-diagonal elements vs the distance $k$ from the diagonal.}
\label{fig:ham}
\end{figure}

\begin{figure}[htb]
\includegraphics[width=0.45\textwidth]{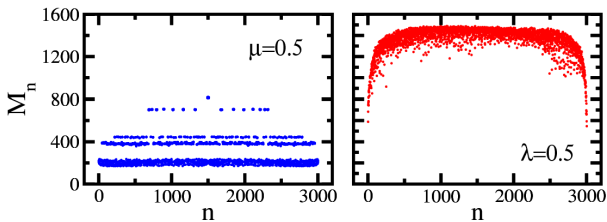}
\caption{(Color online.) Connectivity of each line $n$ for Model 1
(left) and Model 2 (right).}
\label{fig:connect}
\end{figure}

Further details about the matrices may be obtained with the help of Figs.~\ref{fig:ham} and \ref{fig:connect}.

(i) The diagonal elements $H_{nn}$ are shown in the top panels of Fig.~\ref{fig:ham}. Changes are seen as the perturbation increases, especially for Model 2. This indicates that contributions to $H_{nn}$ come not only from the unperturbed part of the Hamiltonians, but also from the perturbation. Also noticeable is an asymmetry between low and high energies, which is enhanced for larger perturbation. For Model 1, larger values of $|H_{nn}|$ are reached for negative energies, while the opposite occurs for Model 2. This imbalance is carried to various other properties of the systems, as will be seen later.

(ii) The bottom panels of Fig.~\ref{fig:ham} show the average values of the absolute values of the off-diagonal elements, $\langle H_{n,n+k} \rangle =[\sum_{n=1}^{D_{L/3}-k} |H_{n,n+k}|]/(D_{L/3}-k)$, vs the distance $k$ from the diagonal. They are significantly smaller than the diagonal elements and decrease slowly as we move away from the diagonal. Thus, even though the Hamiltonians in the site-basis have only NN and NNN couplings, long range (but finite) interactions become present in the mf-basis.

(iii) Figure~\ref{fig:connect} shows the values of the connectivity $M_n$ of each line $n$, that is the number of directly coupled basis vectors in each row.  We present results for $\mu,\lambda =0.5$; they do not change much for larger values of the perturbation. To compute $M_n$, we discard the off-diagonal elements $H_{nm}$ for which $|H_{nm}|<\eta$, where $\eta$ is the variance of the absolute value of all off-diagonal elements. This is done, because the Hamiltonian is initially written in the site-basis and then
numerically transformed into the mf-basis,
which causes all matrix elements to become nonzero.
%; tiny ones corresponding in fact to absence of coupling.

The connectivity for the integrable model is significantly lower than for the chaotic system. For Model 2 in the middle of the spectrum, almost all basis vectors with the same parity are coupled. On average, for the middle of the spectrum, we find
\begin{eqnarray}
&& \mbox{Model 1:} \hspace{0.5 cm} \langle M_n \rangle \sim D_P/4, \nonumber \\
&& \mbox{Model 2:} \hspace{0.5 cm} \langle M_n \rangle \sim D_P.
\label{connect_1_2}
\end{eqnarray}
This confirms that $H_1$ is more sparse than $H_2$, as already observed in Fig.~\ref{fig:structure}. Also in connection to that figure, we see here an interesting structure of separated layers for the values of the connectivity of Model 1, which must be related to its integrability. For Model 2, on the other hand, $M_n$ has a smoother behavior with $n$.

From the Hamiltonian matrix we can estimate also the relative
strength of the perturbation. For this, we compare for each line the average value of the coupling strength $v_n$ with the mean level spacing $d_n$ between directly coupled states.
Taking into account that not all unperturbed states are directly coupled, we define $v_n = \sum_{m\neq n} |H_{nm}|/M_{n}$ and compute the mean level spacing from $d_n=[\varepsilon_n^{max}-\varepsilon_n^{min}]/M_{n}$, where $\varepsilon_n^{max}$ ($\varepsilon_n^{min}$) is the unperturbed energy $H_{mm}$ corresponding to the largest (smallest) $m$ where $H_{nm}\neq0$. Strong perturbation is achieved when $v_n/d_n \gtrsim 1$.

\begin{figure}[htb]
\includegraphics[width=0.45\textwidth]{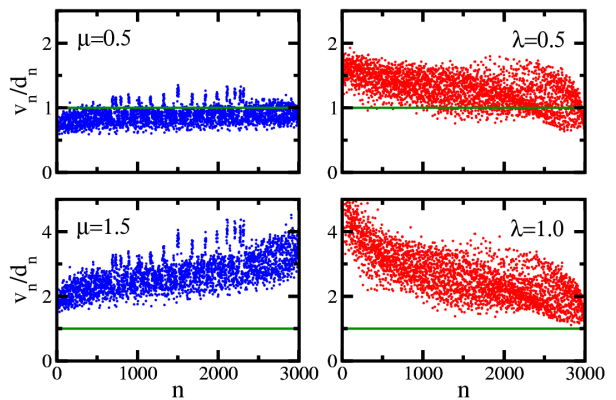}
\caption{(Color online.) Ratio of the average coupling strength $v_n$ to the mean level spacing $d_n$ between directly coupled states for each line $n$ of Model 1 (left panels) and Model 2 (right panels). Horizontal (green) line stands for $v_n/d_n=1$.}
\label{fig:V_df}
\end{figure}

Figure~\ref{fig:V_df} depicts the ratio $v_n/d_n$ for Model 1 (left panels) and Model 2 (right panels). The critical values above which the perturbation becomes strong are approximately $\mu_{cr}\sim 0.5$ and $\lambda_{cr} \sim 0.5$. As we will show later, these estimates coincide with values obtained using the eigenvalues and eigenstates of the systems. Interestingly, the ratio is not flat; it increases with $n$ for Model 1 and decreases with $n$ for Model 2. This is a reflection of the asymmetry of the diagonal elements, already seen in Fig.~\ref{fig:ham}, and it will reappear in Sec.~\ref{IPR_S} when we discuss the level of delocalization of the eigenstates.

\section{Signatures of Quantum Chaos: eigenvalues}
\label{Sec:chaos}

Different quantities exist to identify the crossover from integrability to quantum chaos. Level spacing distribution, level number variance, and rigidity~\cite{MehtaBook,HaakeBook,Guhr1998,ReichlBook}, for example, are associated with the eigenvalues, the first being the most commonly used signature of chaos. In this section, we show briefly some results for the level spacing distribution after having a look at the density of states.

\subsection{Density of states}

\begin{figure}[htb]
\includegraphics[width=0.4\textwidth]{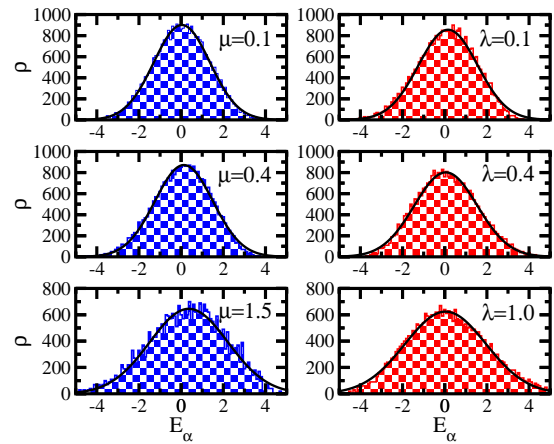}
\caption{(Color online.) Density of states for Model 1 (left panels)
and Model 2 (right panels); bin size = 0.1. The solid (black) line gives the best Gaussian fit:
$\mu=0.1 \rightarrow \langle E \rangle = 0.034, \sigma = 1.330$;
$\mu=0.4 \rightarrow \langle E \rangle = 0.131, \sigma = 1.375$;
$\mu=1.5 \rightarrow \langle E \rangle = 0.363, \sigma = 1.857$;
$\lambda=0.1 \rightarrow \langle E \rangle = 0.157, \sigma = 1.400$;
$\lambda=0.4 \rightarrow \langle E \rangle = 0.051, \sigma = 1.494$;
and
$\lambda=1.5 \rightarrow \langle E \rangle = 0.037, \sigma = 1.920$}
\label{fig:rhoE}
\end{figure}

We denote the eigenvalues of the system by $E_{\alpha}$ and the eigenstates by $|\alpha\rangle$. The density of states $\rho (E_{\alpha})$ for both models are seen in Fig.~\ref{fig:rhoE}. Since the Hilbert space is finite, $\rho (E_{\alpha})$ consists of two parts. On the left side of the spectrum, $\rho (E_{\alpha})$ increases with energy; there the microcanonical
temperature is positive. The right side corresponds to negative temperatures. The point of maximum density of states has infinite temperature.

Independently of the domain, the distributions are very close to
Gaussians, as typical of systems with few-body interactions (two-body in our case)~\cite{French1970,Brody1981}. This is to be contrasted with ensembles of full random matrices, where the density of states is semicircular~\cite{HaakeBook,Guhr1998,ReichlBook}. The fact that the density of states vanishes at very low and very high energies implies that ergodic states are not expected to be found in the edges of the spectrum, even if the system is chaotic. Our analyses of the shapes of the eigenstates, developed in the next section, concentrate thus on the middle of the spectrum.

\subsection{Level spacing distribution}

The analysis of the level spacing distribution requires unfolding the spectrum of each symmetry sector separately. Unfolding the spectrum consists of locally rescaling the energies, so that the mean level density of the new sequence of energies is unity~\cite{HaakeBook,Guhr1998,ReichlBook}. Here, we discard 20\% of the energies located at the edges of the spectrum, where the fluctuations are large, and obtain the cumulative mean level density by fitting the staircase function with a polynomial of degree 15.

\begin{figure}[htb]
\includegraphics[width=0.4\textwidth]{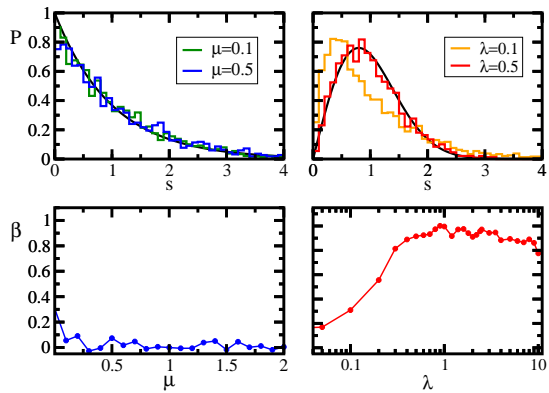}
\caption{(Color online.) Top panels: Level spacing distribution. Bottom panels: Parameter $\beta$ of the Brody distribution vs the perturbation strength. Left panels: Model 1; right panels: Model 2.}
\label{fig:Ps}
\end{figure}

Quantum levels of integrable systems are not prohibited from crossing and the distribution is typically Poissonian,
\[
P_{ P}(s) = \exp(-s),
\]
where $s$ is the normalized level spacing. This is the distribution obtained for Model 1 with any value of $\mu$, as shown in the left top panel of Fig.~\ref{fig:Ps}. In chaotic systems, crossings are avoided and the level spacing distribution is given by the Wigner-Dyson distribution, as predicted by random matrix theory. Ensembles of random matrices with time reversal invariance, the so-called Gaussian Orthogonal Ensembles (GOEs), lead to
\[
P_{ WD}(s) = (\pi s/2)\exp(-\pi s^2/4).
\]
This is the distribution obtained for Model 2 in the chaotic limit, as shown in the right top panel of Fig.~\ref{fig:Ps}. Notice, however, that our systems, contrary to GOEs, have only finite-range-two-body interactions and do not contain random elements. Practically, $P(s)$ is not capable of detecting these differences and the same is expected for other signatures of quantum chaos associated with the energy levels, such as rigidity and level number variance. For an idea of how the results for the level number variance would look like, we refer the reader to Fig.5 in \cite{lea01}, where an equivalent system is considered. More details about the system are found in the properties associated with the eigenstates, as further discussed in the next section.

The parameter $\beta$, used to fit $P(s)$ with the Brody distribution~\cite{Brody1981},
\[
P_B(s) = (\beta +1) b s^{\beta} \exp \left( -b s^{\beta +1} \right), \hspace{0.2 cm}
b= \left[\Gamma \left( \frac{\beta + 2}{\beta +1} \right)\right]^{\beta +1},
\]
can be used to quantify the level of chaoticity of the system reflected by the spectrum statistics. For the integrable Model 1, $\beta$ is close to 0 for any value of $\mu$ (left bottom panel of Fig.~\ref{fig:Ps}), while for Model 2 (right bottom panel), it changes from 0 to 1 as $\lambda$ increases~\cite{noteIntegrable}. The crossover from integrability to chaos is fast and occurs for $\lambda_{cr} \sim 0.5$. This value coincides with the estimate derived from the Hamiltonian matrix in Fig.~\ref{fig:V_df}. It is impressive that the latter procedure, which does not require the diagonalization of the Hamiltonian, can give such satisfactory result.

\section{Signatures of Quantum Chaos: eigenstates}
\label{Sec:EFs}

In this section we explore the features of the eigenstates

\[
|\alpha\rangle=\sum_{n} C^{\alpha}_{n} |n\rangle
\]
written in the mf-basis $|n \rangle $ for both integrable and chaotic regimes. As will soon become clear, more information about the system may be found in the structures of EFs than in the eigenvalues.

\begin{figure}[htb]
\includegraphics[width=0.4\textwidth]{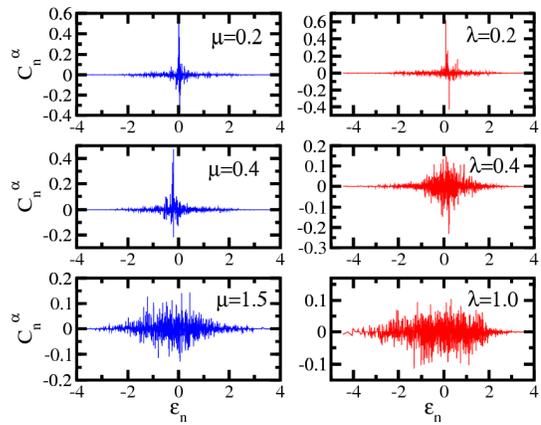}
\caption{(Color online.) Examples of eigenstates from the center of the spectrum for Model 1 (left) and Model 2 (right). They become more extended from top to bottom.}
\label{fig:EF_examples}
\end{figure}

Standard perturbation theory applies when the perturbation is weak, $v_n/d_n \ll 1$. In this limit, the eigenstates are very similar to the mf-basis states, having a very small number of very large components $C^{\alpha}_{n}$. As the perturbation increases, $|\alpha \rangle$ spreads in the unperturbed basis, and the number of principal components,
$N_{pc}$, eventually gets very large. This transition is illustrated in Fig.~\ref{fig:EF_examples}. The eigenstates are shown as a function of the unperturbed energy $\varepsilon_n$ rather than in the basis representation, following the one-to-one correspondence between each unperturbed state $|n \rangle $ and its energy $\varepsilon_n$.

\begin{figure}[htb]
\vskip 0.4 cm
\includegraphics[width=0.4\textwidth]{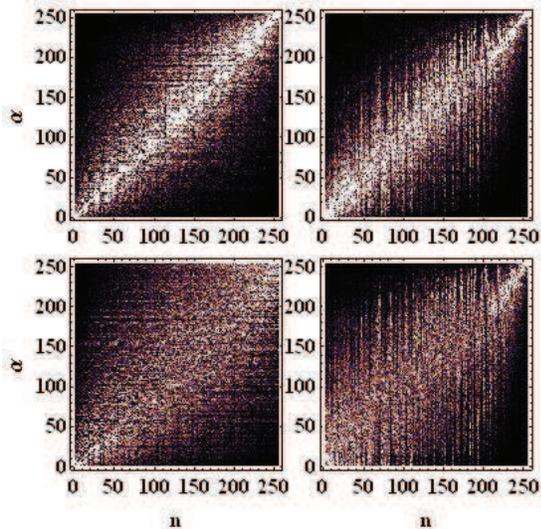}
\caption{(Color online.) Matrix of squared components of the eigenstates for Model 1 (left): $\mu=0.5$ (top) and $\mu=1.5$ (bottom); and Model 2 (right): $\lambda=0.5$ (top) and $\lambda=1.0$ (bottom). Only even states are shown, $L=12$. Light color indicates large values.}
\label{fig:EF_matrix}
\end{figure}

Notice that even for very large perturbation, not all vectors $|n\rangle$ contribute to the eigenstates. The restricted number of participating basis states is a consequence of the finite range of the interactions; only part of the unperturbed states is directly coupled and therefore able to integrate the eigenstates. The limited spread of EFs is clearly seen in Fig.~\ref{fig:EF_matrix}, where the squared amplitudes
%$w_n^{\alpha}=|C^{\alpha}_n|^2$
$|C^{\alpha}_n|^2$ are depicted. In the figure, the basis representation is used. Each horizontal line corresponds to an eigenstate of energy $E_{\alpha}$ in the unperturbed basis. Vertical lines are the unperturbed states with energy $\varepsilon_n$ projected onto the basis of exact states. Light colors represent large $|C^{\alpha}_n|^2$.
%$w_n^{\alpha}$.
The widths of participating states in the vertical and horizontal lines are similar; they are broader in the middle of the spectrum and spread further as the perturbation increases. As $\mu$ and $\lambda$ increase, the differences in magnitude between diagonal and off-diagonal elements become less pronounced. The asymmetry between the edges of the spectrum observed in Figs.~\ref{fig:ham} and \ref{fig:V_df} is seen here again, localization being more enhanced for low energies in Model 1 and for high energies in Model 2 (see bottom panels). Also noticeable is a difference in sparsity between EF and SF depending on the system. For Model 2, just contrary to what was observed for Wigner band random matrix models~\cite{Casati1996}, EFs seem to be more sparse than SFs.

\subsection{Strength function and energy shell}
\label{SFandShell}

In the energy representation, the strength function corresponds to the dependence of $|C^{\alpha}_n|^2$ on the exact energies $E_{\alpha}$ for each fixed unperturbed energy $\varepsilon_n$. It is given by the expression,
\begin{equation}
P_n(E) = \sum_{\alpha} |C^{\alpha}_n|^2 \delta (E-E_{\alpha}),
\end{equation}
where the sum is performed over a small energy window centered at $E$.

For an initial state $|n_0\rangle$, $P_{n_0}$ identifies the energies $E_{\alpha}$ that become available to the state when the perturbation is turned on. The width of SF is therefore associated with the lifetime of $|n_0\rangle $. This is clearly seen by the relation between the probability $W_{n_0}(t)$ for the system to remain in the state and SF, as given by
\begin{eqnarray}
W_{n_0}(t) &=& \left| \langle  n_0 | e^{-i H t}|n_0 \rangle \right|^2 =
\left| \sum_{\alpha} |C^{\alpha}_{n_0}|^2 e^{-i E_{\alpha} t} \right|^2 \nonumber \\
&& \approx \left| \int dE P_{n_0}(E) e^{-iEt} \right|^2,
\label{W0}
\end{eqnarray}
where
\begin{equation}
P_{n_0}(E) = \overline{|C^{\alpha}_{n_0}|^2} \rho(E)
\label{SF_rho}
\end{equation}
is SF after replacing the sum over a large number of eigenstates by an integral, the bar stands for an average in a small energy window, and $\rho(E)$ is the density of exact eigenstates.

An important aspect of SF is the possibility of measuring it experimentally. In nuclear physics, this is done by exciting an unperturbed state and studying its decay. In solid state physics, SF corresponds to the local density of states, since it gives the density of states for an electron on position $|n\rangle $.

\begin{figure}[htb]
\vskip 0.4 cm
\includegraphics[width=0.4\textwidth]{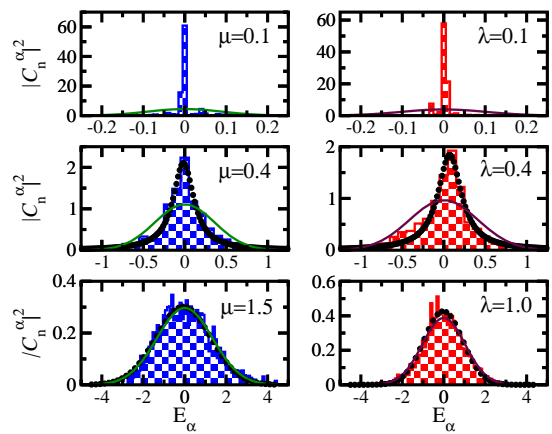}
\caption{(Color online.)  Strength functions for Model 1 (left) and Model 2 (right) obtained by averaging over 5 even unperturbed states in the middle of the spectrum. The average is performed after shifting the center of SFs to zero.
Circles give the fitting curves.
Middle panels: Breit-Wigner with
$\varepsilon + \delta = -0.015$, $\Gamma = 0.302$ (left)
and
$\varepsilon + \delta  = 0.072$, $\Gamma = 0.345 $ (right).
Bottom panels: Gaussian with
$\langle E \rangle  = -0.072$, $\sigma = 1.322$ (left)
and
$\langle E \rangle  = -0.022$, $\sigma = 0.936$ (right).
Solid curves correspond to the Gaussian form of the energy shells with
$\sigma = 0.090$ for $\mu =0.1$;
$\sigma = 0.359$ for $\mu =0.4$;
$\sigma = 1.345$ for $\mu =1.5$;
$\sigma = 0.103$ for $\lambda =0.1$;
$\sigma = 0.412$ for $\lambda =0.4$;
and
$\sigma = 1.029$ for $\lambda =1.0$.}
\label{fig:SF}
\end{figure}

SFs, just like EFs, become more spread as the perturbation increases, as illustrated in Fig.~\ref{fig:SF}. We show with filled curves the average shape of SF for 5 even unperturbed states in the middle of the spectrum. SF starts as a delta function. As the interparticle interactions increase, it acquires first a Breit-Wigner (Lorentzian) shape (middle panels), and eventually becomes Gaussian (bottom panels). This agrees with previous studies of quantum many-body systems~\cite{zele,ZelevinskyRep1996,Flambaum1997b,Flambaum2000}.

(i) According to those studies, the Breit-Wigner function is given by
\begin{equation}
P_n(E) = \frac{1}{2\pi} \frac{\Gamma_n }{(\varepsilon_n + \delta_n - E)^2 +
\left[ \Gamma_n /2 \right]^2},
\label{BW_SF}
\end{equation}
where the width $\Gamma_n $ is given by the Fermi Golden Rule,
\begin{equation}
\Gamma_n   \approx 2\pi \overline{|H_{nm}|^2} \rho_m,
\end{equation}
$\delta_n$ is a correction to the unperturbed energy $\varepsilon_n$ due to the residual interaction, $\overline{|H_{nm}|^2}$ is the mean squared value of nonzero off-diagonal
elements of the Hamiltonian, and $\rho_m$ is the density of basis states $|m\rangle$ directly coupled to the initial state $|n\rangle$ via $H_{nm}$.

(ii) The Gaussian form is
\begin{equation}
P_n(E)=\frac{1}{\sqrt{2 \pi \sigma_n^2}} \exp \left( \frac{-(E-\varepsilon_n)^2}
{2\sigma_n^2} \right),
\label{Gaussian_SF}
\end{equation}
where
\begin{equation}
\sigma_n  = \sqrt{\sum_{m\neq n} |H_{nm}|^2}.
\label{dispersion}
\end{equation}
In the following we will assume that, in the center of the band where maximal chaos is realized,
$\Gamma_n = \Gamma$ and $\sigma_n = \sigma$.

The transition from one shape to the other is determined by the relation between $\Gamma$ and $\sigma$ \cite{FI00}. Equation~(\ref{BW_SF}) holds when the perturbation is small, but non-perturbative, $\Gamma \ll \sigma$, while for
$\Gamma \gtrsim \sigma$, SF becomes close to a Gaussian, as in Eq.~(\ref{Gaussian_SF}).

The maximal shape of SF, as given by Eq.~(\ref{Gaussian_SF}), is reached when the diagonal elements of the Hamiltonian matrix become negligible. In this case, SF coincides with the energy shell. The latter corresponds to the density of states obtained from a matrix filled only with the off-diagonal elements of the perturbation~\cite{Casati1993,Casati1996}. It measures the maximum number of basis states coupled by the perturbation.

We computed the energy shells numerically and verified that they agree very well with the Gaussian functions (\ref{Gaussian_SF}) with dispersion (\ref{dispersion}). The solid lines in Fig.~\ref{fig:SF} represent these functions. As follows from Eq.~(\ref{dispersion}), $\sigma^2$ is obtained without any diagonalization. That expression is derived from the distribution of exact eigenvalues $E_{\alpha}$ for each unperturbed state $|n\rangle$, according to~\cite{Flambaum1997b}
\begin{eqnarray}
\sigma^2 &=& \langle E_{\alpha}^2 \rangle - \langle E_{\alpha} \rangle^2
= \sum_{\alpha}  |C^{\alpha}_{n}|^2 E_{\alpha}^2 - \left( \sum_{\alpha}  |C^{\alpha}_{n}|^2 E_{\alpha} \right)^2
\nonumber \\
&=& \sum_{m} \langle n|H|m \rangle \langle m|H |n\rangle - \varepsilon_n^2
= \sum_{m\neq n} |H_{nm}|^2. \nonumber
\label{sigma_shell}
\end{eqnarray}

As seen in Fig.~\ref{fig:SF}, it is only at large perturbation that SF acquires a Gaussian form and approaches the energy shell. When SF becomes Gaussian with the same width of the energy shell, maximal ergodic filling of the energy shell is realized and a statistical description becomes possible. The agreement between SF and the energy shell  is another way to find the critical values $\mu_{cr}$ and $\lambda_{cr}$. We fitted our numerical data with both functions (circles in Fig.~\ref{fig:SF}) and verified that the transition from Breit-Wigner to Gaussian happens for the same critical values, $\mu_{cr} \, , \lambda_{cr} \approx 0.5$, obtained before from $v_n/d_n$ in Fig.~\ref{fig:V_df} and from the transition to a Wigner-Dyson distribution in the case of Model 2. At large perturbation we then have an excellent agreement between the Gaussian fit and the Gaussian describing the energy shell which depends only on the off-diagonal elements of the Hamiltonian matrices. As seen in the bottom panels, these two curves become practically indistinguishable.

Notice that even at very large perturbation, the width of the energy shell, and thus of the maximal SF, is narrower than the width of the density of states (cf. Fig.~\ref{fig:SF} and Fig.~\ref{fig:rhoE}), especially for Model 2.
This contradicts the equality between $P_n(E)$ and $\rho(E)$ found in previous works~\cite{Flambaum2001a} and may be due to the fact that here the perturbation acts also along the diagonal (such effect is typically removed by considering a renormalized mf-Hamiltonian that takes into account the diagonal contributions of the perturbation).

\subsection{Emergence of chaotic eigenstates}
\label{chaoticEF}

The energy shell determines the maximum fraction of unperturbed states that are accessible to EFs. Therefore, notions of localized ($N_{pc} \sim 1$) or delocalized ($N_{pc}\gg 1$) eigenstates make sense only with respect to the energy shell. When the perturbation is not very strong, large values of $N_{pc}$ may already be found, but in this case EFs are sparse and the components fluctuate significantly. It is only at strong perturbation that the eigenstates can fill the energy shell ergodically, becoming in this way chaotic states~\cite{Casati1993,Casati1996}
and allowing for a statistical description  of the system.
In this limit, the coefficients $C^{\alpha}_{n}$ become random variables from a Gaussian distribution and %$w_n^{\alpha}$
$|C^{\alpha}_{n}|^2$ fluctuate around the envelope defined by the energy shell.

\begin{figure}[htb]
\vskip 0.4 cm
\includegraphics[width=0.4\textwidth]{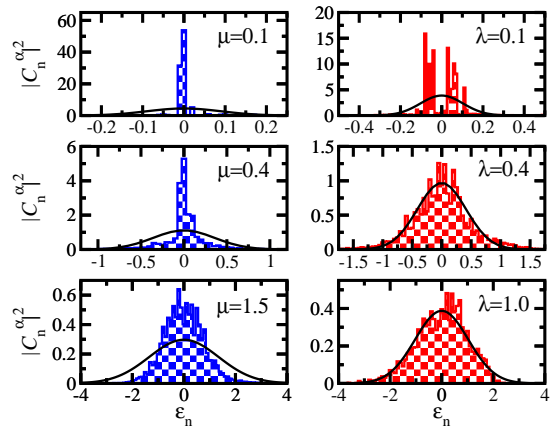}
\caption{(Color online.)  Eigenstates for Model 1 (left) and Model 2 (right) obtained by averaging over 5 even perturbed states in the middle of the spectrum. The average is performed after shifting the center of EFs to zero. They are shown with filled curves. Solid curves correspond to the Gaussian form of the energy shells.}
\label{fig:EF}
\end{figure}

The top panels of Fig.~\ref{fig:EF} show strongly localized states. For Model 2, the EFs are also sparse. The transition to extended states in the energy shell occurs again at the same critical parameters $\mu_{cr} \, , \lambda_{cr} \approx 0.5$, confirming the predictions based on the estimates obtained from $v_n/d_n$ and the Gaussian form of SFs. Notice, however, that EFs from Model 1 never become completely extended, not even for $ \mu =1.5$, although they do fill a large part of the energy shell. We may argue that EFs become chaotic-like, but not truly chaotic. This lack of ergodicity has its roots in the integrability of the system. For Model 2, on the other hand, EFs fill the shell ergodically when the perturbation is strong, being therefore truly chaotic.

Distinctions between integrable and chaotic regimes are thus not captured by SFs, which are ergodic for both models when $\mu$ and $\lambda$ are large. Therefore, ergodicity in SFs implies extended but not necessarily chaotic eigenstates. By comparing EFs and SFs, it becomes evident that even though their structures should be related, since both are derived from $|C^{\alpha}_{n}|^2$, differences do exist. 

\subsection{Delocalization measures}
\label{IPR_S}

Measures quantifying the level of delocalization of individual EFs reveal
further
differences between integrable and nonintegrable models. Overall larger fluctuations appear for the integrable case, which agrees with recent results obtained for bosonic and fermionic
systems~\cite{lea01,lea}.

Delocalization measures \cite{Izrailev1990,ZelevinskyRep1996}, such as the inverse participation ratio (IPR) or the Shannon (information) entropy S, determine the degree of complexity of individual states. For eigenstates in the mf-basis, they are respectively defined as
\begin{equation}
\mbox{IPR}_{\alpha} \equiv \frac{1}{\sum_n |C^{\alpha}_n|^4}
\label{IPR}
\end{equation}
and
\begin{equation}
\mbox{S}_{\alpha} \equiv -\sum_n  |C^{\alpha}_n|^2 \ln |C^{\alpha}_n|^2.
\label{entropyS}
\end{equation}
These quantities measure how much spread the eigenstates are in the unperturbed basis. To quantify the level of
delocalization of the mf-basis vectors with respect to the compound states, we may simply compute the analogous quantities $\mbox{IPR}_n$ and $\mbox{S}_n$, where the sum over $n$ in Eqs.~(\ref{IPR}) and (\ref{entropyS}) are replaced by sums over $\alpha$.

Complete delocalization occurs for GOEs, where the amplitudes $C^{\alpha}_n$ are independent random variables from a Gaussian distribution and the weights $|C^{\alpha}_n|^2$ fluctuate around $1/D$, $D$ being the dimension of the random matrix. The average over the ensemble leads to $\mbox{IPR}_{\text{GOE}}\sim D/3$ and $\mbox{S}_{\text{GOE}} \sim \ln(0.48 D)$ \cite{Izrailev1990,ZelevinskyRep1996}. For the realistic systems considered here, since their eigenstates are confined to energy shells, the values of $\mbox{IPR}$ and $\mbox{S}$ cannot reach those of GOEs~\cite{Kota_IPR}.

\begin{figure}[htb]
\includegraphics[width=0.45\textwidth]{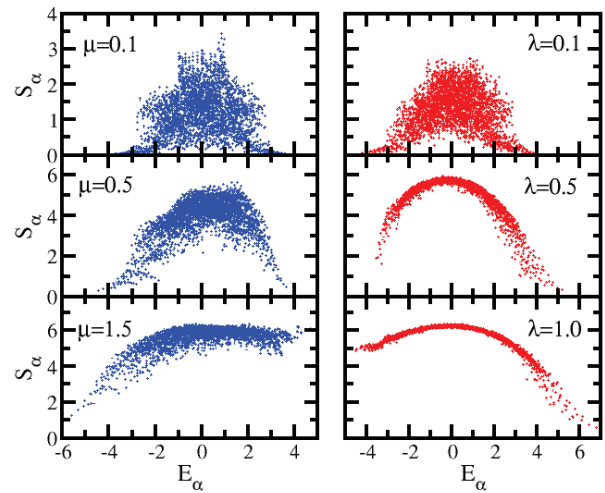}
\caption{(Color online.) Shannon entropy for all eigenstates written in the mf-basis for Model 1 (left) and Model 2 (right).}
\label{fig:S_EF}
\end{figure}

Figure~\ref{fig:S_EF} shows $\mbox{S}$ for the eigenstates of Model 1 (left panels) and Model 2 (right panels).
As expected from the shape of the density of states (see Fig.~\ref{fig:rhoE}), strong mixing occurs in the middle of the spectrum, $\mbox{S}$ being smaller at the edges. Interestingly however, large values of $\mbox{S}$ are still found at the borders when the perturbation is very strong. For Model 1 this happens at high energies and for Model 2 at low energies; following the same asymmetry verified before (cf. Figs.~\ref{fig:ham} and \ref{fig:V_df}).

As the perturbation increases from top to bottom panels in Fig.~\ref{fig:S_EF}, the values of $\mbox{S}$ increase and the fluctuations decrease for both models. However, this reduction is much more significant for Model 2. The smooth behavior of $\mbox{S}$ in the chaotic limit (bottom right panel) indicates that the structure of eigenstates close in energy becomes statistically very similar. This fact has suggested a close relationship between chaos and the viability of thermalization~\cite{Deutsch1991,Srednicki1994}, as numerically explored in \cite{lea01,lea}.

Differences between integrable and chaotic regimes, as verified in the behavior of $\mbox{S}$ and in the spreading of EFs in the energy shell (see Fig.~\ref{fig:EF}), appear to have their origins in the results for the connectivity shown in Fig.~\ref{fig:connect}. The separated values of $M_n$ seen in the integrable system must lead to EFs with different levels of delocalization, even when close in energy. This causes larger fluctuations in the values of $\mbox{S}$. For Model 2, $M_n$'s are similar for nearby states leading to the smooth behavior of $\mbox{S}$ in the bottom right panel of Fig.~\ref{fig:S_EF}.

\begin{figure}[htb]
\includegraphics[width=0.45\textwidth]{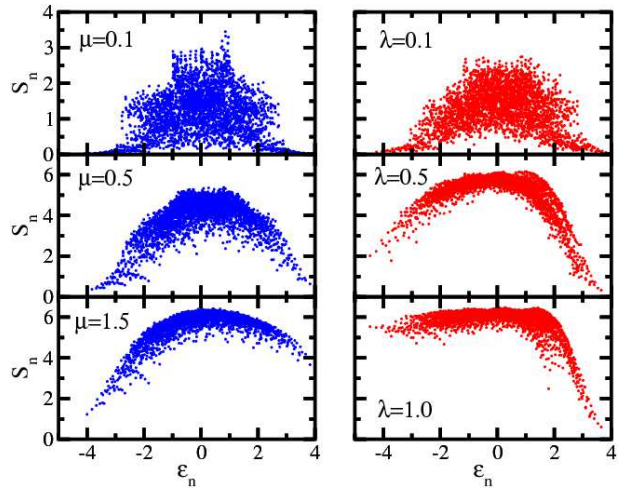}
\caption{(Color online.) Shannon entropy for the strength functions written in the basis of the eigenstates for Model 1 (left) and Model 2 (right).}
\label{fig:S_SF}
\end{figure}

The level of delocalization of SFs for the basis states written in terms of the eigenstates also increases with the perturbation, while the fluctuations decrease, as shown in Fig.~\ref{fig:S_SF}. Here however, the width of the fluctuations are very similar for both models. This reinforces our previous statement that the SF cannot capture differences between the two models, showing comparable behavior for both integrable and nonintegrable systems.

\subsection{Overlap between neighboring eigenstates}

We define a new signature of chaos referred to as the overlap between the probability distributions of neighboring eigenstates $|\alpha \rangle$ and $|\alpha' \rangle$,
% and defined as
%
\begin{equation}
\Omega_{\alpha, \alpha'} \equiv \sum_n  |C^{\alpha}_n|^2 |C^{\alpha'}_n|^2.
\end{equation}
It corresponds to an alternative way to capture the transition to chaos by measuring how much similar the components of neighboring states are.

\begin{figure}[htb]
\includegraphics[width=0.45\textwidth]{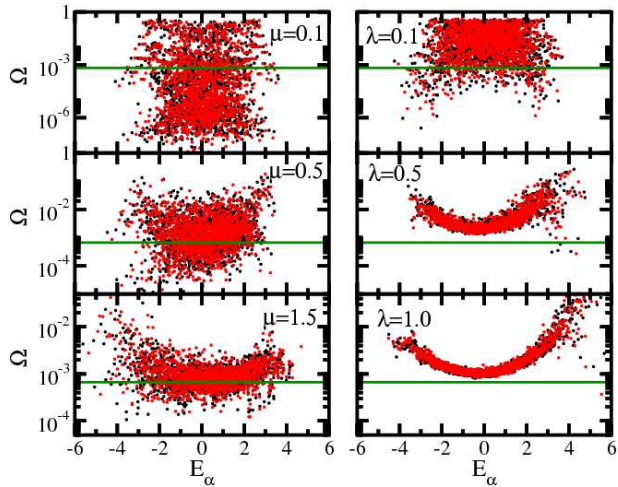}
\caption{(Color online.) Overlaps of neighboring eigenstates for Model 1 (left panels) and Model 2 (right panels). Dark (Black) and light (red) points indicate eigenstates of even or odd parity. Horizontal (green) lines indicate the GOE prediction $\Omega = 1/D$.}
\label{fig:overlap}
\end{figure}

For GOEs, since all eigenstates are simply normalized pseudo-random vectors, one has  $\Omega \sim 1/D$. These states are completely delocalized and statistically very similar. For the models studied here, the results are presented in Fig.~\ref{fig:overlap} and described below.

(i) In the limit of localized eigenstates, large fluctuations are seen. Since there are few contributing components, we find neighboring states where the probabilities $|C^{\alpha}_n|^2$ are nonzero and approximately the same for the same basis vectors $|n\rangle$, but we have also pairs where the effective basis vectors do not match. There are very correlated states leading to large overlaps, and there are also
uncorrelated states leading to values of $\Omega$ below the threshold from GOEs, the values reached by Model 1 being significantly lower than for Model 2.

(ii) As the perturbation increases, and the number of principal components becomes large, the maximum values of $\Omega$ decrease for both regimes, especially in the middle of the spectrum where the mixing is stronger. The fluctuations in the values of the overlaps also decrease, especially for Model 2. For the latter, a smooth behavior with energy, similar to that obtained  for the Shannon entropy for EFs, is achieved.

(iii) Notice that in the limit of strong perturbation, only Model 2 does not cross the GOE threshold. In the integrable model, since EFs do not fill the energy shell completely, we may still find neighboring states that are statistically very different. At the edges of the spectrum, the overlaps tend to be larger, since there are more correlations due to finite effects.

\section{Time evolution of the Shannon entropy: statistical relaxation}
\label{Sec:Shannon}

We now study the quench dynamics of the system by focusing on the time evolution of the Shannon entropy for initial states corresponding to unperturbed vectors selected from the middle of the spectrum. For an initial state $|n_0\rangle $, the entropy in the mf-basis is given by
\begin{equation}
S_{n_0}(t)=-\sum_{n=1}^{D_P}  W_{n} (t) \ln \left[ W_{n} (t) \right]
\label{entropy}
\end{equation}
where
\[
W_{n} (t) = \langle n| e^{-iH t} |n_0 \rangle =\left| \sum_{\alpha} C^{\alpha}_n C^{\alpha*}_{n_0} e^{-iE_{\alpha}t} \right|^2
\]
is the probability for the initial state $|n_0 \rangle$ to be found in the state $|n\rangle$.

Numerical data are shown in Fig.~\ref{fig:dynamics}. To reduce fluctuations, an average is performed over 5 initial even basis states excited in a narrow energy window in the middle of the spectrum. In the limit of strong interaction, the results for both the chaotic and the integrable models agree very well with analytical expressions previously found in the context of two-body-random ensembles \cite{Flambaum2001b}. These expressions can be derived when the shape of SF is known, being either Breit-Wigner or Gaussian.

\subsection{Analytical expressions}

We reproduce here the steps of the cascade model considered in Ref.~\cite{Flambaum2001b} to obtain an analytical expression for the time dependence of the entropy.

For very short times, $t\ll \Gamma/\sigma^2$, it has been shown that the probability for the system to remain in the initial state $|n_0\rangle $ is~\cite{FlambaumAust,Flambaum2001a}
\begin{equation}
W_{n_0}(t)  \approx \exp(-\sigma^2 t^2).
\label{short_time}
\end{equation}
For very long times the probability becomes
\begin{equation}
W_{n_0}(t) \approx \exp(-\Gamma t),
\label{long_time}
\end{equation}
which means that the decay rate from the initial state is determined by
\begin{equation}
\frac{d W_{n_0}}{dt} = -\Gamma W_{n_0}.
\label{dW0}
\end{equation}
Given the two-body interaction, $|n_0\rangle $ spreads first into $N_1$ states directly coupled to it. This set is referred to as the first class of states.
Subsequently, states from the first class populate those directly coupled to them,
the $N_2$ basis states from the second class. The process continues successively like this as in a cascade~\cite{Alts}. The number of states in the $k$-th class  is then
\begin{equation}
N_k = M_k \ldots M_1 M_{n_0} \approx M_{n_0}^k,
\label{calM}
\end{equation}
where $M_k$ is the connectivity associated with the basis states of
the  $k$-th class.
This implies that the number of states of one class
is larger than the number in the previous
class, which justifies neglecting the probability of return to a
previous class and allows us to write, for $k>1$
\begin{equation}
\frac{d {\cal C}_k}{dt} = \Gamma {\cal C}_{k-1} - \Gamma {\cal C}_k,
\label{dWk}
\end{equation}
where ${\cal C}_k$ is the probability for the system to be in the $k$-th class and ${\cal C}_0=W_{n_0}$.
The first term on the right-hand side is the flux from the previous class
and the second term is the decay of the  $k$-th class.

\begin{figure}[htb]
\includegraphics[width=0.4\textwidth]{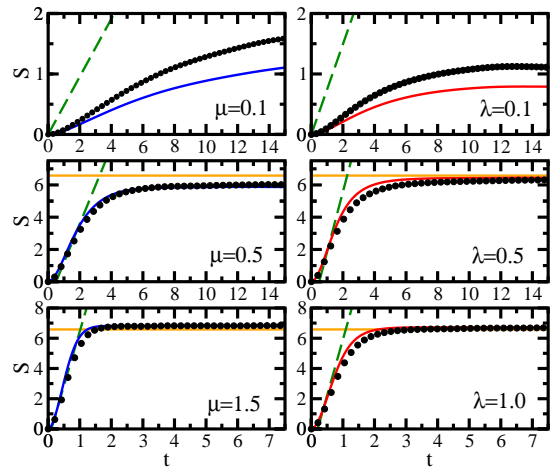}
\caption{(Color online) Shannon entropy vs time for Model 1 (left) and Model 2 (right). Circles stand for numerical data,
dashed lines show the linear dependence (\ref{linear}), and solid curves correspond to Eq.~(\ref{S_analytical}). The horizontal (orange) solid lines represent the value of $\mbox{S}_{\text{GOE}} \sim 6.58$.}
\label{fig:dynamics}
\end{figure}

The solution of  Eq.~(\ref{dWk}) is
\begin{equation}
{\cal C}_k = \frac{(\Gamma t)^k}{k!} e^{-\Gamma t}.
\label{classes}
\end{equation}
Since each $k$  class contains several basis states, ${\cal C}_k \approx N_k W_{n} $. Assuming an infinite number of classes, Eq.~(\ref{entropy})  becomes
\begin{eqnarray}
S_{n_0}(t) &\approx& - \sum_{k=0}^{\infty} {\cal C}_k \ln \left( \frac{{\cal C}_k}{N_k} \right) \nonumber \\
&=& \Gamma t \ln M_{n_0} + \Gamma t - e^{-\Gamma t}
\sum_{k=0}^{\infty} \frac{(\Gamma t)^k}{k!} \ln \frac{(\Gamma t)^k}{k!}. \nonumber
\end{eqnarray}
The last terms on the right-hand side of this equation are smaller than the first term, so they may be neglected, leading to a simple linear time dependence of the Shannon entropy,
\begin{equation}
S_{n_0}(t) \approx \Gamma t \ln M_{n_0}.
\end{equation}
In the limit of strong perturbation, where $\Gamma \gtrsim \sigma$ and SF is described by a Gaussian, we can write the entropy as
\begin{equation}
S_{n_0}(t) \approx \sigma_{n_0} t \ln M_{n_0}.
\label{linear}
\end{equation}
Note that Eq.~(\ref{linear}) depends only on the elements of the Hamiltonian matrix. Yet, as seen in Fig.~\ref{fig:dynamics}, it reproduces very well the linear increase of the entropy for both models in the regime where the eigenstates become delocalized in the energy shell.

To find an expression that describes the dynamics of the system at both short and long times, Eq.~(\ref{short_time}) needs to be taken into account. In Ref.~\cite{Flambaum2001b}, the following expression was proposed,
\begin{eqnarray}
S_{n_0}(t) = &-& W_{n_0} (t) \ln W_{n_0}(t) \nonumber \\
&-&
\left[ 1- W_{n_0} (t)\right] \ln \left( \frac{1- W_{n_0} (t)}{N_{pc}}\right),
\label{S_analytical}
\end{eqnarray}
where $N_{pc}$ is the total number of states inside the energy shell, that is the limiting value of the entropy after relaxation. In the results shown in Fig.~\ref{fig:dynamics}, we obtained $N_{pc}$ numerically from $N_{pc}=\langle e^S \rangle$, where the average $\langle. \rangle$ is performed over a long time interval after the entropy saturates, $t \in [100,200]$.

Equation~(\ref{S_analytical}) is a good approximation when the total number of classes is small, $n_c \sim 1$. This is indeed the case for Models 1 and 2. The effective number of classes in the cascade model can be obtained from
\begin{equation}
M^{n_c} = D_P,
\end{equation}
which, following Eq.~(\ref{connect_1_2}), gives $n_c\sim1.2$ for Model 1 and $n_c \sim 1$ for Model 2.

In the regime of strong perturbation, Eq.~(\ref{S_analytical}) captures all stages of the evolution: the initial quadratic growth, as given by perturbation theory; the linear behavior; and the final saturation. For small perturbation, the agreement with Eq.~(\ref{S_analytical}) is poor. Notice, however,  that the perturbation here was not sufficiently small to show oscillations as in~\cite{Smerzi}

The main aspects of the statistical relaxation process are then the linear growth of S followed by its saturation to a value close to that of a GOE: $\mbox{S}_{\text{GOE}} \sim \ln(0.48 D)$. In the limit of strong interaction, this is the behavior of the chaotic system and, to a very good approximation, also the behavior of the integrable model. This suggests that chaoticity is not essential for the emergence of statistical relaxation. The fact that EFs of both models in the limit of large interaction show significant filling of the energy shell indicates that the existence of extended eigenstates is a sufficient condition for relaxation. However, to reach a final statement, further numerical and analytical studies of one- and two-body observables are necessary.

\section{Conclusion}
\label{Sec:conclusion}

We studied static and dynamic properties of two systems of interacting spins 1/2. Model 1 is integrable for any value of the perturbation and Model 2 can transition to chaos. The analysis of the Hamiltonian matrices, later combined with studies of spectrum statistics and structures of eigenstates and strength functions, suggested that aspects of the intricate behavior of complex systems can be anticipated even before diagonalization.

It was shown that strength functions and eigenstates delocalize as the perturbation increases, being, however, always restricted to the energy shell. In the limit of strong perturbation, strength functions of both models in the middle of the spectrum become Gaussian and coincide with the energy shell. In the case of eigenstates, the same occurs only for the chaotic model. For the integrable system, the eigenstates become much spread, but do not fill the energy shell completely.

We verified that the lack of ergodicity of the eigenstates for the integrable model is reflected in larger fluctuations of delocalization measures and of the overlaps between neighboring eigenstates. The degree of overlaps between neighboring eigenstates may be considered as a new signature of chaos. The transition to chaos occurs when the values of the overlaps becomes inversely proportional to the dimension of the Hilbert space.

We also studied the time evolution of the Shannon entropy for initial states corresponding to mean-field basis vectors. Knowledge of the shape of the strength functions allowed us to describe the quench dynamics with analytical expressions originally developed and tested for systems with two-body-random interactions. They agreed very well with our numerics. The linear growth of the entropy was also well described by an expression involving parameters obtained from the analysis of the Hamiltonian matrices before diagonalization.

Our results indicate that the relaxation process is very similar for integrable and nonintegrable systems, provided the eigenstates are extended in the energy shell. On the other hand, we have seen that after saturation the fluctuations of the entropy in the integrable domain are slightly larger than for the chaotic system, as observed also in \cite{rigol,Balachandran2010} in the context of observables.

An issue that deserves further investigation concerns the fluctuations of static and dynamic properties. A careful analysis of how they reduce with the number of particles and how the results compare for both regimes is very important for further developments of the problem of thermalization in isolated systems.

\begin{acknowledgments}

L.F.S. was supported by the NSF under grant DMR-1147430. F.B. was supported by Regione Lombardia and Consorzio Interuniversitario Lombardo per L'Elaborazione Automatica through a Laboratory for Interdisciplinary Advanced Simulation Initiative grant  (2010) [http://lisa.cilea.it]. He also acknowledges support from Universit\'a Cattolica Grant No. D.2.2 2010. F.M.I. acknowledges support from Consejo Nacional de Ciencia y Tecnolg\'ia Grant No. N-161665 and thanks Yeshiva University for the hospitality during his stay.
\end{acknowledgments}


\begin{thebibliography}{72}
\bibitem[{not()}]{noteIntegrable}
\bibinfo{note}{There are in fact two transitions. The system is integrable in
  both extremes, when the NNN couplings are nonexistent ($\lambda=0$) and also
  when the NN couplings become negligible ($\lambda \rightarrow \infty$), but
  this second limit is unrealistic.}
\end{thebibliography}
\end{document}